\documentclass[aps,prx,amsmath,amssymb,twocolumn,superscriptaddress]{revtex4-1}

\usepackage{graphicx}
\usepackage{natbib}
\usepackage{color}

\begin{document}

\title{Giant pressure-enhancement of multiferroicity in CuBr$_2$}

\author{J.~S.~Zhang}
\thanks{These authors contributed equally to this study.}
\affiliation{Mathematics and Physics Department, North China Electric Power
University, Beijing 102206, China}
\affiliation{Department of Physics, and Beijing Key Laboratory of
Opto-electronic Functional Materials $\&$ Micro-nano Devices, Renmin
University of China, Beijing 100872, China}
\author{Yiqi~Xie}
\thanks{These authors contributed equally to this study.}
\affiliation{International Center for Quantum Materials, School of Physics,
Peking University, Beijing 100871, China}
\author{X. Q. Liu}
\affiliation{International Center for Quantum Materials, School of Physics,
Peking University, Beijing 100871, China}
\author{A. Razpopov}
\affiliation{Institut f\"ur Theoretische Physik, Goethe-Universit\"at
Frankfurt, 60438 Frankfurt am Main, Germany}
\author{V. Borisov}
\affiliation{Institut f\"ur Theoretische Physik, Goethe-Universit\"at
Frankfurt, 60438 Frankfurt am Main, Germany}
\author{C. Wang}
\affiliation{International Center for Quantum Materials, School of Physics,
Peking University, Beijing 100871, China}
\author{J. P. Sun}
\affiliation{Beijing National Laboratory for Condensed Matter Physics and
Institute of Physics, Chinese Academy of Sciences, Beijing 100190, China}
\affiliation{School of Physical Sciences, University of Chinese Academy of
Sciences, Beijing 100190, China}
\author{Y. Cui}
\affiliation{Department of Physics, and Beijing Key Laboratory of
Opto-electronic Functional Materials $\&$ Micro-nano Devices, Renmin
University of China, Beijing 100872, China}
\author{J. C. Wang}
\affiliation{Department of Physics, and Beijing Key Laboratory of
Opto-electronic Functional Materials $\&$ Micro-nano Devices, Renmin
University, Beijing 100872, China}
\author{X. Ren}
\affiliation{International Center for Quantum Materials, School of Physics,
Peking University, Beijing 100871, China}
\author{Hongshan Deng}
\affiliation{Center for High Pressure Science and Technology Advanced
Research (HPSTAR), Beijing 100193, China}
\author{Xia Yin}
\affiliation{Center for High Pressure Science and Technology Advanced
Research (HPSTAR), Beijing 100193, China}
\author{Yang Ding}
\affiliation{Center for High Pressure Science and Technology Advanced
Research (HPSTAR), Beijing 100193, China}
\author{Yuan~Li}
\email[]{yuan.li@pku.edu.cn}
\affiliation{International Center for Quantum Materials, School of Physics,
Peking University, Beijing 100871, China}
\affiliation{Collaborative Innovation Center of Quantum Matter, Beijing 100871,
China}
\author{J. G. Cheng}
\email[]{jgcheng@iphy.ac.cn}
\affiliation{Beijing National Laboratory for Condensed Matter Physics and
Institute of Physics, Chinese Academy of Sciences, Beijing 100190, China}
\affiliation{School of Physical Sciences, University of Chinese Academy of
Sciences, Beijing 100190, China}
\affiliation{Songshan Lake Materials Laboratory, Dongguan, Guangdong 523808,
China}
\author{Ji~Feng}
\affiliation{International Center for Quantum Materials, School of Physics,
Peking University, Beijing 100871, China}
\affiliation{Collaborative Innovation Center of Quantum Matter, Beijing
100871, China}
\author{R. Valent\'{\i}}
\affiliation{Institut f\"ur Theoretische Physik, Goethe-Universit\"at
Frankfurt, 60438 Frankfurt am Main, Germany}
\author{B. Normand}
\affiliation{Neutrons and Muons Research Division, Paul Scherrer Institute,
CH-5232 Villigen-PSI, Switzerland}
\author{Weiqiang~Yu}
\email[]{wqyu\_phy@ruc.edu.cn}
\affiliation{Department of Physics, and Beijing Key Laboratory of
Opto-electronic Functional Materials $\&$ Micro-nano Devices, Renmin
University of China, Beijing 100872, China}


\begin{abstract}
Type-II multiferroic materials, in which ferroelectric polarization is
induced by inversion non-symmetric magnetic order, promise new and highly
efficient multifunctional applications based on the mutual control of
magnetic and electric properties. Although this phenomenon has to date been
limited to low temperatures, here we report a giant pressure-dependence of
the multiferroic critical temperature in CuBr$_2$. At 4.5 GPa, $T_\mathrm{C}$
is enhanced from 73.5 to 162 K, to our knowledge the highest value
yet reported for a non-oxide type-II multiferroic. This growth shows no
sign of saturating and the dielectric loss remains small under these high
pressures. We establish the structure under pressure and demonstrate a 60\%
increase in the two-magnon Raman energy scale up to 3.6 GPa. First-principles
structural and magnetic energy calculations provide a quantitative explanation
in terms of dramatically pressure-enhanced interactions between CuBr$_2$
chains. These large, pressure-tuned magnetic interactions motivate structural
control in cuprous halides as a route to applied high-temperature
multiferroicity.
\end{abstract}

\maketitle

\section{Introduction}
\label{si}

The search for application-suitable multiferroics \cite{SpaldinScience2005,
RameshNatMater2007,SpaldinPhysT2010} has advanced significantly over the
last decade in both type-I and type-II materials \cite{MaAM2011,ZhaoNC2014,
OrtegaJPCM2015,FiebigNRM2016,SeixasPRL2016,HuangPRL2018}. Type-I multiferroics
\cite{KhomskiiPhys2009} have independent magnetic and ferroelectric transitions
\cite{CatalanAM2009,JainNPJQM2016}, meaning that even when both transition
temperatures are high, the magnetoelectric coupling, and hence the scope for
mutual control, is usually weak. The physics of most type-II multiferroics
\cite{KhomskiiPhys2009,KimuraNature2003,CheongNatMater2007,ZhangJACS2018}
involves frustrating magnetic interactions that give rise to a spiral magnetic
order \cite{TokuraAdvMater2010}, which immediately generates a ferroelectric
polarization by the inverse Dzyaloshinskii-Moriya mechanism
\cite{KatsuraPRL2005,SergienkoPRB2006,MostovoyPRL2006,DongPRB2008,
XiangPRL2008}. However, an intrinsic drawback of magnetic frustration
is that it suppresses the onset of long-range order, and hence most
currently available type-II multiferroics operate only at low temperatures
\cite{CheongNatMater2007}.

A generic route to higher operating temperatures in type-II multiferroics
is to increase the strength of the magnetic interactions. This can, in
principle, be achieved through structural alterations, for which perhaps
the cleanest method is an applied pressure \cite{HaumontPRB2009,VargaPRL2009,
RuizPRB2012,RocquefelteNC2013,AoyamaNC2014}. Pressure, broadly construed to
include chemical pressure and substrate pressure, acts to increase electronic
hybridization without introducing disorder. In the most minimal model for a
magnetic insulator, the antiferromagnetic (AF) exchange interaction is given
by $J = 4t^2/U$, where $t$ is the orbital hybridization and $U$ the on-site
Coulomb repulsion. However, excessive $t$ risks driving the system metallic,
thus losing its magnetic and ferroelectric properties. The most scope for
achieving large $J$ values is offered by large initial values of both $t$
and $U$, making the spin-1/2 Cu$^{2+}$ ion particularly promising in view of
its often strong on-site correlations and significant orbital hybridization
with ligands. It is not a coincidence that complex copper oxides become
high-temperature superconductors after charge-carriers are introduced into
the Mott-insulating parent compounds \cite{LeeRMP2006}, or that CuO is a
type-II multiferroic with the equal highest transition temperature ($T_\mathrm{C}
\simeq 230$~K) known to date \cite{KimuraNatMater2008}.

\begin{figure}
\includegraphics[width=8.5cm]{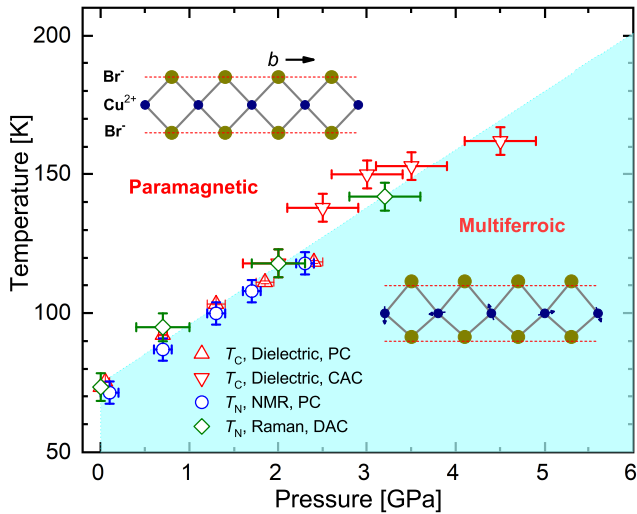}
\caption{\label{pd} Phase diagram of CuBr$_2$: $T_\mathrm{C}(P)$ and
$T_\mathrm{N}(P)$ determined from dielectric, NMR, and Raman scattering
measurements. A piston cell (PC), a cubic anvil cell (CAC), and a diamond
anvil cell (DAC) were used for different pressure ranges and measurements
as specified. The upper inset shows a schematic representation of the chain
structure in the high-temperature paramagnetic phase; the lower shows a
representation of the chain structure and magnetism in the low-temperature
[$T < T_\mathrm{N}(P)$] multiferroic phase, where the magnetic order (arrows)
is helical in the chain direction, breaking inversion symmetry, and the
ferroelectricity is caused by small displacements of the Br$^{-}$ ions
(exaggerated for illustration).}
\end{figure}

CuBr$_2$ is a non-oxide type-II multiferroic material with a CdI$_2$-type
monoclinic structure \cite{ZhaoAdvMater2012}. The structural units are
CuBr$_4$ squares, which form edge-sharing chains in the $b$ direction (insets,
Fig.~\ref{pd}). These chains have a $C$-centered stacking in the $a$ direction
and coincidentally form nearly coplanar units in the $b(a$$+$$c)$ plane. Early
first-principles calculations~\cite{LeePRB2012} of the magnetic interactions
indicated that the dominant coupling [$J_5$ in Fig.~\ref{structure}(a)] is
that between next-nearest-neighbor Cu$^{2+}$ ions within the chains, which
is AF. Other strong interactions are expected to be the ferromagnetic (FM)
nearest-neighbor in-chain bond ($J_1$) and the AF coupling between sites in
coplanar chains ($J_7$); additional weak interactions were suggested to be
responsible for the formation of long-ranged three-dimensional(3D) magnetic
order. Frustration between $J_1$ and $J_5$ suggests spiral order along the
chains, with the spin rotation angle given classically by $\theta =
\cos^{-1} (- J_1/4 J_5)$, which approaches 90$^\circ$ when $J_5$ significantly
exceeds $|J_1|$. At ambient pressure, a spiral magnetic order does indeed
develop below $T_\mathrm{N} = 73.5$ K, with an incommensurate propagation
wave vector (1, 0.2350, 0.5) \cite{ZhaoAdvMater2012,LeePRB2012} fully
consistent with the expected $\theta$. Spontaneous electric polarization
is detected immediately below $T_\mathrm{N}$, defining a rather high
ferroelectric transition temperature \cite{ZhaoAdvMater2012}; while
$T_\mathrm{C} = T_\mathrm{N}$ by definition in a type-II multiferroic, below
we distinguish between the two according to our method of experimental
detection. Similar properties have been found in the isostructural compound
CuCl$_2$, albeit at considerably lower temperatures \cite{SekiPRB2010}.

Here we report our investigation of ferroelectricity and magnetism in CuBr$_2$
under hydrostatic pressure. By combined dielectric-constant, nuclear magnetic
resonance (NMR), Raman-scattering, and x-ray diffraction (XRD) measurements in
three different types of pressure cell, we have established the ($P,T$) phase
diagram up to pressures of 4.5 GPa. As shown in Fig.~\ref{pd}, we find a rapid
and massive pressure-driven enhancement of the multiferroic transition
temperature. Density functional theory (DFT) calculations based on the XRD
structure establish that the equally rapid rise of the two-magnon Raman energy
is a consequence of the dramatic pressure-sensitivity of the Cu-Br-Br-Cu $J_7$
interaction, while $T_\mathrm{N}$ and hence $T_\mathrm{C}$ are driven primarily
by the inter-plane coupling (most strongly by $J_2$). There is no evidence for
saturation of this behavior up to the largest pressures studied, where the
material remains highly insulating, confirming that there is plenty of room
at the top for pressure tuning of $T_\mathrm{C}$ in CuBr$_2$.

The structure of this article is as follows. In Sec.~\ref{smm} we present
details of our sample and experimental methods, and summarize our theoretical
analysis. In Sec.~\ref{sdm} we present the results of our dielectric
measurements in two different types of pressure cell. Section \ref{ssm}
shows analogous results for the magnetic properties obtained by NMR and
Raman scattering. In Sec.~\ref{sst} we analyze our high-pressure structural
measurements by detailed first-principles calculations of electronic and
magnetic energies, from which we explain the pressure evolution of all of
the magnetic interactions governing the behavior of CuBr$_2$. Section \ref{sdc}
contains a discussion of our results, some perspective on the prospects they
offer for applicable multiferroics, and a brief summary.

\section{Material and Methods}
\label{smm}

Large single crystals of CuBr$_2$ were grown by slow evaporation from aqueous
solutions \cite{ZhaoAdvMater2012}. Because ferroelectric transitions usually
cause sudden changes in the dielectric constant, we attached two copper-plate
electrodes to the opposing $ab$ faces of a plate-like crystal to form a
capacitor with the electric field applied perpendicular to the $ab$ plane.
Measurements of the capacitance as a function of temperature ($T$), pressure
($P$), and magnetic field ($B$) were used  to indicate when a spontaneous
ferroelectric polarization had developed. A crystal with dimensions
4$\times$1.5$\times$0.4 mm$^3$ was used for dielectric measurements in
a piston cell (PC) and one with dimensions 0.7$\times$0.7$\times$0.2 mm$^3$
in a cubic anvil cell (CAC).
The softness and propensity to chemical dissolution of the crystal meant
that dielectric measurements above 2.4 GPa were possible only in the CAC,
but not yet in anvil cells with smaller sample spaces as reported in the
literature \cite{Honda2014}; the larger error bars on the corresponding data
points in Fig.~\ref{pd} reflect the complexity of these measurements. The
sample and copper plates were connected using an inert epoxy and suspended
in a Teflon capsule filled with Daphne oil as the pressure-transmitting
medium. The pressure was calibrated at room temperature by monitoring the
characteristic resistance changes of Bismuth (Bi). The capacitance
was measured by an Agilent 4263B LCR meter with an excitation level
of 1.0 V at 100 kHz.

The zero-field $^{81}$Br ($I = 3/2$) NMR spectra were measured by the
spin-echo method using $\pi/2$-$\tau$-$\pi$ sequences, where $\pi/2$
and $\pi$ denote RF pulses with respective time durations of 0.5
and 0.8 $\mu$s, and the time interval is $\tau = 6$ $\mu$s. The pressure
was calibrated using the $^{63}$Cu nuclear quadrupole resonance (NQR)
frequency of Cu$_2$O in the pressure cell \cite{RRSI1992}. The electronic
Raman scattering measurements were performed in a confocal backscattering
geometry using the 632.8 nm line of a He-Ne laser. The low-$T$ and
high-$P$ conditions were realized using an Almax easyLab diamond anvil
cell (DAC) integrated into a Janis ST-500 optical cryostat, with Argon
as the pressure-transmitting medium. The pressure was calibrated by the
fluorescence line of a ruby sphere loaded together with sample inside the
DAC. The high-pressure XRD experiments were performed at pressures up to
14.73 GPa at beamline 16 BM-D of the HPCAT sector at the Advanced Photon
Source (APS), Argonne National Laboratory, using a Mao-type symmetric DAC.
CuBr$_2$ powder samples and ruby chips were loaded into the sample chamber
with Neon gas as the pressure-transmission medium. Diffraction patterns were
recorded on a MAR345 image plate and integrated by DIOPTAS software.

First-principles calculations of the structural and magnetic properties of
CuBr$_2$ were carried out using density functional theory with the electronic
correlations for the Cu $3d$ states included at the mean-field level within
the Generalized Gradient Approximation (GGA)+$U$ approach. In the first
(structural) step, the lattice parameters at selected pressure values
were taken from experiment and the internal positions of the Br ions
were optimized using the Vienna {\it ab initio} simulation package (VASP)
\cite{rkf}. In the second (magnetic) step, the magnetic interaction parameters
[Fig.~\ref{structure}(c)] were estimated from the structures at each pressure
by computing the total magnetic energies in 27 different spin configurations
using the all-electron full-potential local-orbital (FPLO) basis code
\cite{rke} and then performing a total-energy mapping to a pure Heisenberg
model with 9 different bilinear parameters, $J_i$ [Fig.~\ref{structure}(a)].

\begin{figure}
\includegraphics[width=8.5cm]{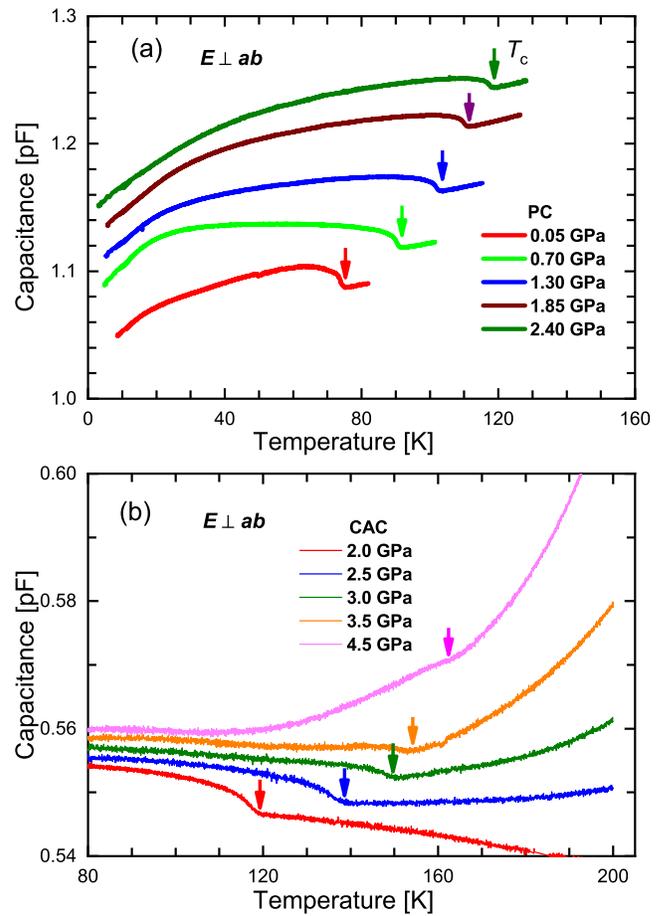}
\caption{\label{dielectric} Dielectric response under pressure, deduced
from the capacitance between two copper plates attached to single crystals
(see text), shown as a function of temperature. Measurements were performed
with (a) PC and (b) CAC apparatus to reach pressures up to 4.5 GPa. The kink
features (arrows) indicate the ferroelectric transitions.}
\end{figure}

\begin{figure}
\includegraphics[width=8.5cm]{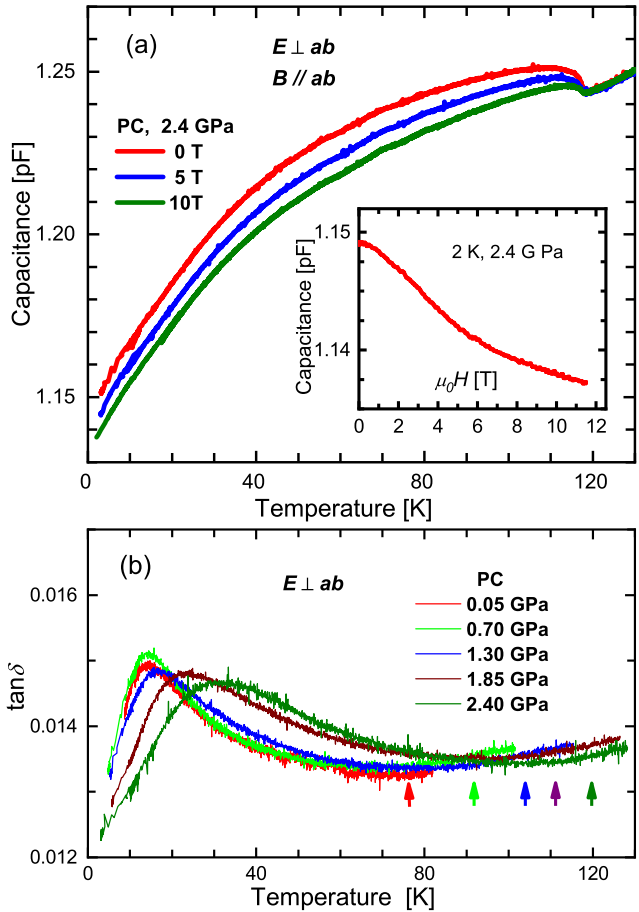}
\caption{\label{dielectric3} Magnetoelectric coupling and dielectric loss.
(a) Capacitance at 2.4 GPa, measured under an external magnetic field, $B$,
applied perpendicular to the electric field, $E$. The inset shows the
field-dependence of the capacitance measured at 2 K and 2.4 GPa.
(b) Dielectric loss, $\tan \delta$, as a function of temperature at
different pressures. $\tan\delta$ shows almost no variation with the
excitation level or frequency.}
\end{figure}

\section{High-pressure dielectric measurements}
\label{sdm}

The dielectric constant is extracted from the capacitance between two
electrodes attached to the $ab$ surfaces of a single crystal, as described
in Sec.~\ref{smm}. Because the sample dimensions change under pressure, we
present the capacitance rather than the dielectric constant. Results from
measurements in the PC with no applied magnetic field are shown
in Fig.~\ref{dielectric}(a) and in the CAC in
Fig.~\ref{dielectric}(b). At $P = 0.05$ GPa, the capacitance at 80 K is 1.07 pF,
which gives a dielectric constant $\varepsilon_\mathrm{r} \simeq 8.1$, close to
the value reported previously at ambient pressure \cite{ZhaoAdvMater2012}.
The onset of ferroelectricity is shown by a sudden increase in capacitance
on cooling below $T_\mathrm{C} \simeq 75$ K, which is slightly higher than the
ambient-pressure value, $T_\mathrm{C} = 73.5$ K. The capacitance decreases
monotonically with further cooling, because of reduced charge fluctuations,
and increases with rising pressure, as might be expected on compression
(reduced inter-layer separations). The remarkable feature of these data is
the dramatic rise of $T_\mathrm{C}$ to 118.5 K at 2.4 GPa in the PC, and further
to 162~K at 4.5~GPa in the CAC (Fig.~\ref{pd}). The latter $T_\mathrm{C}$
represents a 120\% increase over the ambient-pressure value, or an average
growth rate $d T_\mathrm{C} / dP \approx 19.7$ K/GPa. Equally surprisingly,
$T_\mathrm{C}$ continues to rise nearly linearly, with no evidence at 4.5~GPa
for a saturation of the effect.

To verify the presence of a magnetoelectric coupling, we applied an external
magnetic field in the $ab$ plane in our PC measurements. This is expected
to distort the spiral magnetic structure and hence to affect the ferroelectric
properties. As shown in Fig.~\ref{dielectric3}(a), the capacitance at 2.4 GPa
in fields $B = \mu_0 H = 0$, 5, and 10 T is constant above $T_\mathrm{C} =
118.5$ K, and so is $T_\mathrm{C}$ itself. However, the magnitudes of both the
capacitance and the capacitance anomaly increase monotonically with decreasing
field, providing direct evidence both for a significant magnetoelectric
coupling and for magnetic-order-induced (i.e.~type-II) ferroelectricity
\cite{ZhaoAdvMater2012}.

\begin{figure}
\includegraphics[width=8.5cm]{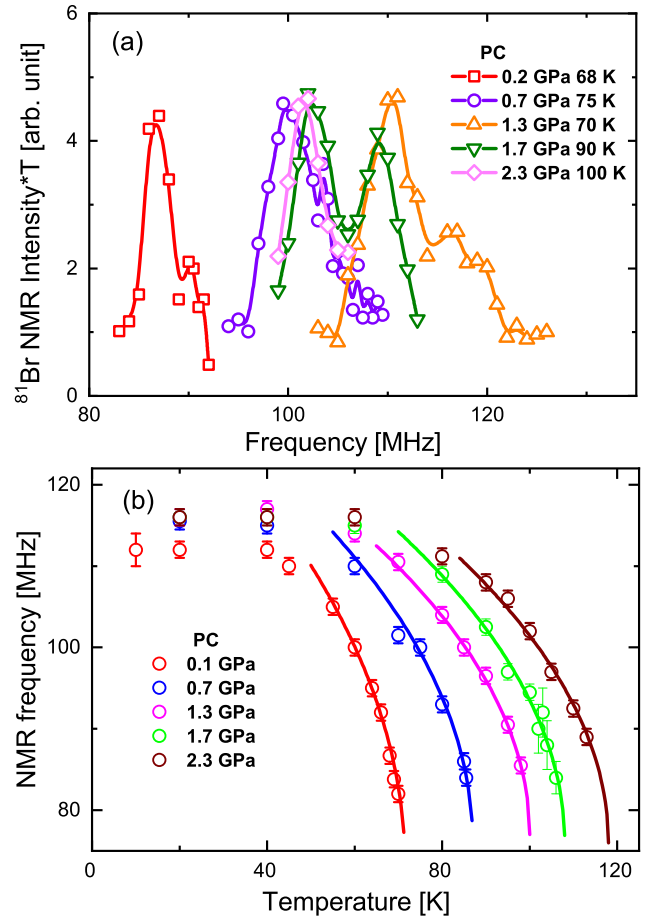}
\caption{\label{spectra} NMR measurements under pressure. (a) Zero-field
$^{81}$Br NMR spectrum, shown in the conventional form of the product of
spin-echo intensity and temperature as a function  of frequency, measured
in a PC at a selection of fixed pressures and temperatures. (b) Zero-field
$^{81}$Br NMR frequency, $f(P,T)$, which drops sharply as the temperature
approaches $T_\mathrm{N}(P)$; solid lines show fits to the form $f(P,T) =
f_0(P,T_\mathrm{N}) + a(P)(T_\mathrm{N} - T)^{1/2}$.}
\end{figure}

The dielectric loss, $\tan\delta$, is an important figure of
merit for the practical application of ferroelectric materials. In
Fig.~\ref{dielectric3}(b) we observe that $\tan\delta = 0.013 \pm 0.001$
above $T_\mathrm{C}$ at all pressures reached in the PC; this value is
again consistent with ambient-pressure data \cite{ZhaoAdvMater2012}. At
all pressures, $\tan\delta$ increases weakly when the sample is cooled
below $T_\mathrm{C}$, forming a broad low-$T$ peak whose center scales with
$T_\mathrm{C}$. Although we do not fully understand the origin of this
feature, one possibility is that the spiral spin configuration continues
to fluctuate until the sample is cooled substantially below $T_\mathrm{C}$,
allowing for a dissipation of electrical energy into the spin system through
the magnetoelectric coupling. These very small values of $\tan\delta$ at all
pressures nonetheless constitute an extremely low dielectric loss, reflecting
both the strongly insulating nature of CuBr$_2$, at least up to 2.5 GPa, and
the considerable potential for device applications.

\begin{figure}
\includegraphics[width=8.5cm]{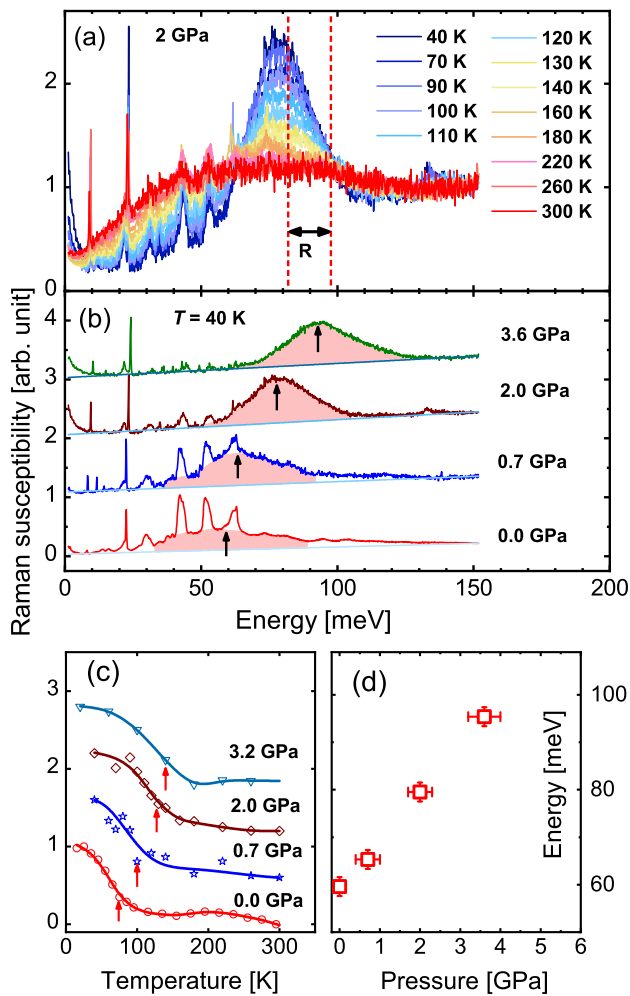}
\caption{\label{raman} Raman scattering measurements under pressure.
(a) Raman spectra measured at 2.0 GPa over the full range of temperatures.
``R'' denotes the frequency window for the averaged Raman susceptibility.
(b) Raman spectra obtained at 40 K under different pressures. Shaded areas
indicate the two-magnon response discussed in the text and arrows indicate
its characteristic central energy. Data are offset vertically for clarity.
(c) Raman susceptibility averaged over an energy interval located directly
above the two-magnon peak center, as shown in panel (a). Arrows indicate
the approximate location of $T_\mathrm{N}(P)$ estimated from a 20\% increase
in the signal on cooling.
(d) Central two-magnon Raman energy shown as a function of pressure.}
\end{figure}

\section{High-pressure NMR and Raman scattering measurements}
\label{ssm}

We have probed the magnetic system by zero-field NMR and Raman-scattering
measurements performed over the same range of pressures as our dielectric
measurements. The $^{81}$Br NMR spectra at all temperatures and pressures,
shown in Fig.~\ref{spectra}(a) as the $T$-weighted spin-echo intensity
as a function of frequency, have one clearly identifiable $I_z = 1/2
\leftrightarrow 3/2$ peak \cite{WangRQ} whose position moves systematically
with both $P$ and $T$. We focus on the resonance frequency, $f(P,T)$, of
this peak at each pressure and display its temperature-dependence in
Fig.~\ref{spectra}(b). Below $T_\mathrm{N}$, $f(P,T)$ has two additive
contributions, one due to the electric-field gradient (EFG), $f_0 (P,T)$,
and one from the static local hyperfine field, $A_{\rm hf} \langle S \rangle$
(where $A_{\rm hf}$ is the hyperfine coupling constant and $\langle S \rangle$
the average magnetic moment). The $f_0$ term is expected to change very little
with temperature at low $T$, and we find it to increase only rather weakly
with pressure. Because the second contribution is proportional to the magnetic
order parameter, $f(P,T)$ decreases sharply as $T \rightarrow T_\mathrm{N}(P)$;
a fit to the form $f(P,T) = f_0(P) + a(P) (T_\mathrm{N}(P) - T)^{1/2}$ at each
$P$ (solid lines) allows us to deduce the values $T_\mathrm{N} (P)$ up to 2.3
GPa shown in Fig.~\ref{pd}. We comment that $f(P,T)$ tends to saturate below
40 K [Fig.~\ref{spectra}(b)], with $f(P,T) - f_0(P) = 39$ MHz both at $P =
0.1$ GPa and at 2.3 GPa, confirming that the ordered moment changes very
little with pressure (under the assumption that $A_{\rm hf}$ does not change
with $P$).

The Raman susceptibility is obtained by dividing the recorded scattered
photon intensities by the Bose factor. Its dominant feature is the
``two-magnon'' excitation \cite{FleuryPR1968,LiYuan2017}, which we show
in Fig.~\ref{raman}(a) for all temperatures at a fixed pressure of 2.0 GPa.
While the sharp peaks are phonons, the two-magnon response is a very broad
peak that in ordered quantum magnets bears little resemblance to the density
of states of 3D spin waves \cite{Devereaux2007}, and persists in the
paramagnetic phase due to short-range magnetic correlations. Here we observe
that this broad peak sharpens at low temperatures to a form quite similar to
the well-characterized cuprate response (of Cu$^{2+}$ spins in a planar quantum
magnet) \cite{Chelwani2018}. Figure \ref{raman}(b) shows this peak (shaded
area) for a fixed low temperature of 40 K at several selected pressures. It
is clear that the two-magnon energy scale increases rapidly under pressure,
rising by 60\% from ambient pressure to 3.6 GPa [Fig.~\ref{raman}(d)].
Despite the complexities inherent to an accurate modelling of the two-magnon
response, it is safe to conclude that the relevant magnetic interactions in
the system are enhanced massively by the effects of hydrostatic pressure.

A subsidiary piece of information may be extracted from the $T$-dependence
of the two-magnon peak intensity, based on the empirical connection between
$T_\mathrm{N}$ and the intensity increase on cooling at ambient pressure
\cite{LiYuan2017}. In Fig.~\ref{raman}(c), we average the Raman susceptibility
over a fixed-percentage energy range [Fig.~\ref{raman}(a)] located slightly
above the central energy of the two-magnon peak (in order to avoid
multi-phonon scattering processes that overlap with the electronic
signal at lower energies). This analysis allows us to extract values
for $T_\mathrm{N}(P)$, marked by the arrows in Fig.~\ref{raman}(c), which
again are fully consistent with the values of $T_\mathrm{C}(P)$ shown in
Fig.~\ref{pd}.

\section{High-pressure structural analysis and DFT calculations}
\label{sst}

\begin{figure*}
\includegraphics[width=16cm]{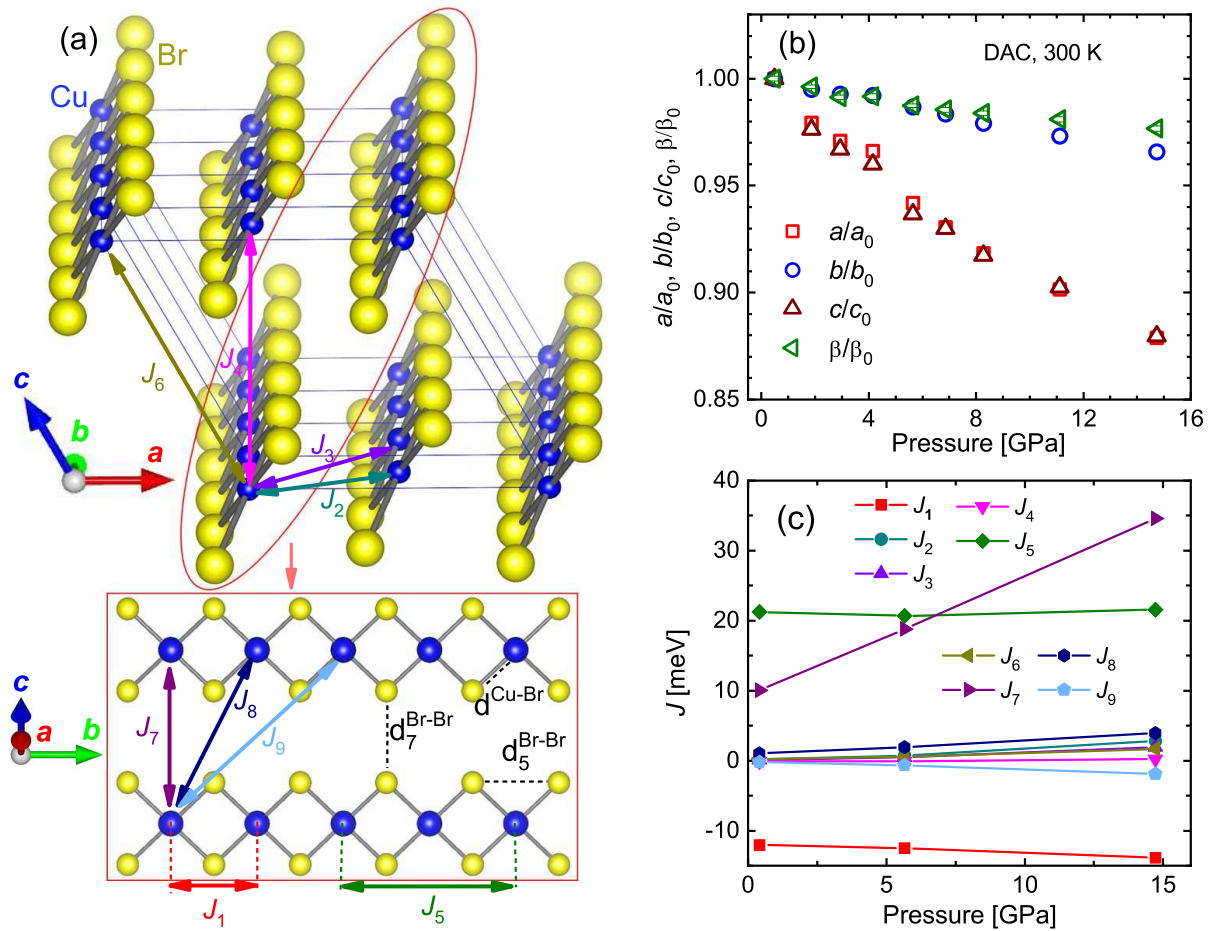}
\caption{\label{structure} Crystal structure and magnetic interactions
under pressure. (a) Crystal structure of CuBr$_2$ viewed nearly parallel
to the $b$ axis. The inset displays the structure of two chains lying almost
precisely in the $b(a$$+$$c)$ plane. The nine Cu-Cu superexchange
interactions computed under pressure are indicated. (b) Relative change
of the lattice parameters $a$, $b$, and $c$, as well as of the angle $\beta$,
determined by XRD at 300~K in a DAC and shown as functions of pressure. The
lattice remains in a monoclinic structure at all pressures. The DAC
measurements are normalized to the base-pressure dimensions obtained at
$P = 0.49$ GPa, $a_0 = 7.09$ \r{A}, $b_0 = 3.47$ \r{A}, $c_0 = 6.91$
\r{A}, and $\beta_0 = 119.3^\circ$. (c) Pressure-dependence of the magnetic
interactions, $J_i$, deduced from the DFT calculations.}
\end{figure*}

\begin{table}[b]
\begin{tabular}{c||c|c|c|c}
$\; P$ [GPa]$\;$ & $\; a $ [\r{A}]$\;$ & $ b $ [\r{A}]$\;$ & $\; c$ [\r{A}]
$\;$ & $\; {\beta}$ [$^\circ$]$\;$ \\
\hline
0.49	& $\;$ 7.09 $\;$ & $\;$ 3.47 $\;$ & $\;$ 6.91 $\;$ & $\;$ 119.3 \\
\hline
1.87	& 6.95	& 3.45	& 6.75	& 118.8 \\
\hline
2.94	& 6.89	& 3.44	& 6.69	& 118.2 \\
\hline
4.15	& 6.85	& 3.44	& 6.64	& 118.3 \\
\hline
5.65	& 6.68	& 3.42	& 6.48	& 117.8 \\
\hline
6.86	& 6.60	& 3.41	& 6.43	& 117.5 \\
\hline
8.28	& 6.52	& 3.40	& 6.34	& 117.3 \\
\hline
11.12	& 6.40	& 3.38	& 6.24	& 117.0 \\
\hline
14.73	& 6.23	& 3.35	& 6.08	& 116.5
\end{tabular}
\caption{Lattice parameters $a$, $b$, and $c$, and monoclinic angle $\beta$,
obtained from the high-pressure XRD data.}
\label{tab1}
\end{table}

Our data in Fig.~\ref{dielectric} and \ref{dielectric3} confirm that
CuBr$_2$ retains qualitatively the same type-II multiferroic properties at
all pressures below 5 GPa. To explain the giant pressure-sensitivity of
$T_\mathrm{C}$ in a quantitative manner, we investigate the structure of
CuBr$_2$ by high-pressure XRD measurements and corresponding DFT
calculations. XRD was performed up to 15 GPa using synchrotron radiation
at the Advanced Photon Source, as outlined in Sec.~\ref{smm}, and we quote
the lattice parameters over the full pressure range in Table~\ref{tab1}.
The key features of our results are first that the monoclinic structure is
preserved for all pressures and second that, as represented graphically in
Fig.~\ref{structure}(b), and as may be expected to lowest order, the chain
units remain rather rigid. There are only small relative changes to the
$b$-axis dimension and the angle $\beta$, whereas the $a$ and $c$ lattice
parameters, which correspond to the chains being compressed together,
change by approximately 12\%. We comment that, although one might expect
the ionic displacement associated with the ferroelectric transition (inset
Fig.~\ref{pd}) to lift the symmetry and interfere with the magnetic
interactions, this value turns out to be truly vanishingly small (from
the pyroelectic current \cite{ZhaoAdvMater2012} one may estimate it to be
0.4 fm on each Br$^-$ ion) and hence plays no role in the structure or
magnetism of the low-$T$ phase.

Our first-principles calculations under pressure are a two-step process as
described in Sec.~\ref{smm}. First we perform a structural optimization
at selected experimental pressure values by fixing the lattice parameters
to those of the corresponding XRD measurements and relaxing the internal
ionic positions within the ``GGA+U'' approach. In this type of calculation,
reliable results are obtained using a $\vec{k}$-point mesh of size
10$\times$10$\times$10, a plane-wave cut-off energy of 800 eV, and
representative Cu-ion correlation parameters $U_{\rm VASP} = 10$ eV and
$J_{\rm H} = 1$ eV; the crystal structures are relaxed until the calculated
ionic forces fall below the threshold $10^{-3}$ eV/\r{A}. For correlated
systems, it is in general necessary to include the spin degrees of freedom
of the transition-metal cation to ensure reliable structural predictions that,
however, are quite insensitive to the actual magnetic order; here a FM order
was imposed, which resulted in a total magnetic moment of $1.0\,\mu_\mathrm{B}$
per formula unit.

For the second step, of interpreting the magnetic properties of CuBr$_2$, we
identify $i = 9$ significant superexchange paths, $\{J_i\}$, as shown in
Fig.~\ref{structure}(a). These span the three spatial dimensions of the system
and their values can be expected to determine the physics of the $b$-axis
chains, the $b(a$$+$$c)$ planes, and 3D magnetic ordering. We then compute
the total energies of 27 different magnetic configurations and map these to
a Heisenberg model with interaction parameters $J_i$. The magnetic-energy
calculations used two different supercells, with dimensions 2$\times$1$\times$2
and 1$\times$4$\times$2, and respective $k$-mesh sizes 12$\times$12$\times$12
and 7$\times$7$\times$7. Electronic correlations were modelled using the
GGA+$U$ functional with $U = 6$ eV and $J_\mathrm{H} = 1$ eV; it is this (FPLO)
value of $U$ which has a direct influence on the energy scale of the magnetic
interactions. The fit to the spin model was performed by a least-squares
regression analysis of the overdetermined system of 27 equations with 11
unknowns (9 superexchange parameters and 2 non-magnetic contributions to
the total energy, one for each supercell). Calculated $J_i$ values for
three representative pressures taken from the XRD study, namely 0.49,
5.65, and 14.73 GPa, are shown in Fig.~\ref{structure}(c). The mean-square
total-energy deviation between the \textit{ab~initio} calculation and the
spin model was 0.06~meV/Cu for $P = 0.49$ GPa, 0.16~meV/Cu for $P = 5.65$
GPa, and 0.52~meV/Cu for $P = 14.73$ GPa, indicating the reliability of the
spin model at all pressures.

\begin{table}[b]
\begin{tabular}{c||c|c|c}
$\; P$ [GPa]$\;$ & $\; d^{\rm {Cu-Br}}$ [\r{A}]$\;$ & $\; d^{\rm{Br-Br}}_5$
[\r{A}]$\;$ & $\; d^{\rm {Br-Br}}_7$ [\r{A}]$\;$ \\
\hline
0.49  &  2.440  &  3.468  &  3.651 \\
\hline
5.65  &  2.421  &  3.423  &  3.380 \\
\hline
14.73  &  2.375  &  3.350  &  3.123
\end{tabular}
\caption{Cu-Br distance, $d^{\rm {Cu-Br}}$, and Br-Br distances, respectively
$d^{\rm {Br-Br}}_5$ and $d^{\rm {Br-Br}}_7$ for the $J_5$ and $J_7$ paths represented
in Fig.~\ref{structure}(a), computed by DFT for pressures of 0.49, 5.65,
and 14.73 GPa.}
\label{tab}
\end{table}

Considering first the chain units, clearly $J_1$ and $J_5$ change rather
little with pressure, which to lowest order may be expected from the small
changes to the $b$-axis lattice parameter [Fig.~\ref{structure}(b)]. In more
detail, the FM \cite{Kanamori1959} Cu-Br-Cu $J_1$ interaction is often very
sensitive to the bond angle, but here this is found to change by less
than 0.5$^\circ$ in a regime close to its optimal value \cite{Oeckler2000}.
While the AF $J_5$ bond is a Cu-Br-Br-Cu ``super-superexchange'' path that
also depends on the bond angle, this remains largely fixed by the rigidity
of the chains.

The dominant physics of the system occurs in the $b(a$$+$$c)$ plane due
to $J_7$, which increases from 10 to 18 meV up to 5.65 GPa and then to 35 meV
at 14.73 GPa. This giant enhancement actually changes the nature of the
planar magnetism from $b$-axis-dominated at ambient pressure to spatially
isotropic at 5 GPa to ($a$$+$$c$)-axis-dominated at 15 GPa; however, in the
absence of significant frustration it has no effect on the $b$-axis spiral
order. The huge rise of $J_7$ under pressure may be understood completely
from the fact that it is also a Cu-Br-Br-Cu path, with the same geometry as
$J_5$ [Fig.~\ref{structure}(a)], and while the Cu-Br distance and angle are
strongly constrained in the chain units, the Br-Br bond in the ($a$$+$$c$)
direction takes up most of the unit-cell compression. As Table \ref{tab}
makes clear, it shrinks from being 0.18 \r{A} longer than the comparable
distance in $J_5$ at 0.49 GPa to 0.23 \r{A} shorter at 14.73 GPa.

Physically, these three interactions create the dominant energy scales in the
magnon dispersion, and $J_7$ would account completely for the rapid pressure
enhancement observed in the two-magnon Raman signal [Fig.~\ref{raman}(b)
and Fig.~\ref{raman}(d)]. Our absolute parameter values are controlled by
the effective $U$ in the calculations, but in CuBr$_2$ it is difficult to
obtain an experimental benchmark due to sample decomposition issues in the
measurement of the high-temperature susceptibility \cite{ZhaoAdvMater2012} and
theoretical issues in interpreting the two-magnon Raman energy (Sec.~\ref{ssm}).
Thus we do not attempt to fit $U$, and simply use a value (6 eV) typical for
insulating inorganic Cu systems. Still, $J_1$, $J_5$, and $J_7$ span only two
spatial dimensions, and to discuss the 3D magnetic order it is necessary to
consider the inter-plane interactions. We find that the second-shortest path
in the system, $J_2$, which creates a zig-zag interchain network in the $ab$
plane, also rises by a factor of 5 from 0.49 to 5.65 GPa (and a further factor
of 4 to 14.73 GPa). Similar rises can also be found in the slightly weaker
$J_3$ and $J_6$ interactions. These results, which are easy to justify by
considering the pressure-induced changes to interchain spin density in the
$ab$ and $bc$ planes, account for the steep rise in $T_\mathrm{N}$,
and hence in $T_\mathrm{C}$, over the pressure range of Fig.~\ref{pd}. It is
clear from our XRD measurements and DFT calculations that this $T_\mathrm{C}$
enhancement can continue to far higher pressures, where $J_7$ will also play
an increasing role in raising $T_\mathrm{N}$, with no intervening structural
transition. These results raise the prospect of room-temperature
multiferroicity in suitably strained CuBr$_2$.

\section{Discussion and Conclusion}
\label{sdc}

Figure~\ref{pd} shows that the two intrinsically linked characteristic
temperatures, $T_\mathrm{C}$ and $T_\mathrm{N}$, as measured by a range of
probes and in a number of different pressure cells, rise strongly with
pressure. Figure~\ref{raman}(d) shows a proportionally similar and
equally quasi-linear rise in the central energy scale determined by
two-magnon Raman scattering. To our knowledge, our maximal $T_\mathrm{C}$
of 162 K, achieved at 4.5 GPa in a CAC, is unprecedentedly high for a
non-oxide type-II multiferroic. Further, although it remains below that
of some oxide type-II multiferroics, such as CuO and YBaCuFeO$_5$
($T_\mathrm{C} \approx 230$ K \cite{KimuraNatMater2008,Morin2015}), many of
these suffer from higher dielectric loss due to their semiconducting nature
\cite{KimuraNatMater2008,RMnO3KatsufujiPRB2001,LiVCuO4PRB2008}. The
persistence of low dielectric loss in CuBr$_2$ under pressure, despite the
increase in orbital hybridization that should move the system towards
metallicity, constitutes a major advantage for electronic applications.

We stress that the characteristic magnetic energy scales in CuBr$_2$, reflected
in the energy of the two-magnon peak, are much higher than $T_\mathrm{N}$. This
indicates that both frustration and dimensionality effects play a strong role
in suppressing $T_\mathrm{N}$ at ambient pressure, and that the effect of
pressure is to reduce both. Indeed, our DFT calculations demonstrate that
the primary change is caused by the interchain ($a$$+$$c$)-axis coupling, $J_7$,
which enhances the 2D nature and makes chain frustration less energetically
relevant. This said, it is important to note that neither the rising $J_7$ nor
any of the other pressure-enhanced interactions has a significant effect on
the existence of the in-chain frustration, which creates the helical $b$-axis
spin state required for type-II multiferroicity. Beyond $J_7$, we have shown
that the interchain $ab$-plane coupling, $J_2$, plays the leading role in
making the system 3D and hence governs the value of $T_\mathrm{N}$; despite
being very low at ambient pressure, its high pressure-sensitivity causes
the strong rise of $T_\mathrm{N}$ whose lower end we have characterized in
the present work. We comment that such massive pressure effects on magnetism
are known in Cu-based metal-organic materials \cite{Wehinger2018}, due to
a combination of soft structures and highly directional ligand paths, but are
uncommon in inorganic Cu systems and to date unknown in multiferroic ones.

In summary, we have demonstrated how strongly the magnetic interactions in
CuBr$_2$ are changed by pressure, and how this makes it possible to effect a
giant enhancement of the multiferroic $T_\mathrm{C}$ using any available methods
for structural control. Dielectric investigations of CuBr$_2$ at pressures
higher than our current limit of 4.5 GPa are certainly required. Alternatively,
different methods of structural tuning, including chemical pressure
\cite{OrganicJerome1991} and epitaxial stress \cite{MuduliJPCP2007},
also affect the magnetism of low-dimensional systems in ways similar to a
hydrostatic pressure. Thin-film growth with epitaxial stress applied along
the $a$- or $c$-direction, by the choice of a suitable substrate, should be
a particularly valuable route to higher $T_\mathrm{C}$ values in CuBr$_2$. We
conclude by stressing once again that the pressures we have investigated
remain far from saturating the $T_\mathrm{C}$ increase in CuBr$_2$, and that
they seem not to impair any of the significant magnetoelectric coupling,
the dielectric loss, or the insulating properties of the material, all
of which present major technical advantages for application purposes.

\begin{acknowledgments}
We thank R. Kremer for helpful discussions.
Work at Renmin University of China was supported by the Ministry of Science
and Technology of China (MOSTC) under Grant No.~2016YFA0300504, the National
Natural Science Foundation of China (NSFC) under Grant No.~51872328 and the
Fundamental Research Funds for the Central Universities and the Research Funds
of Renmin University of China under Grant No.~15XNLQ07. Work at Peking
University was supported by the NSFC (Grants No.~11888101, No.~11874069, 
and No.~11725415)
and the National Basic Research Program of China under Grants
No.~2018YFA0305602. JGC is supported by the MOSTC (Grant
No.~2018YFA0305700), NSFC (Grants No.~11574377, 11834016, and 11874400) and
the Chinese Academy of Sciences (Grants No.~XDB25000000 and QYZDB-SSW-SLH013).
The XRD measurements were performed at sector 16 BM-D (HPCAT) of the Advanced
Photon Source, a U.S. Department of Energy (DoE) Office of Science user
facility operated by Argonne National Laboratory (ANL) under Contract
No.~DE-AC02-06CH11357. HPCAT operations were supported by the DoE-NNSA
under Award No.~DE-NA0001974, with partial instrumentation funding by
the NSF. {\it Ab initio} calculations were supported by the Deutsche
Forschungsgemeinschaft (DFG) through Grant No.~VA117/15-1 with computer
time provided by the Centre for Scientific Computing (CSC) in Frankfurt.
\end{acknowledgments}

\end{document}